
\def\cl{\centerline}
\def\sk{\vskip12pt}
\input harvmac


\Title{{ \vbox{ \hbox{Imperial TP/91-92/32} \hbox{QMW 92-15}
                    \hbox{hep-th/9209095} } } }
{The Phase Space of the Wess-Zumino-Witten Model}
\authorfont
\cl{G. Papadopoulos\foot{Address from Oct. 1st 1992: Dept. Mathematics,
        Kings College, London WC2R 2LS.}}
\sk
{\it\cl{Department of Physics }
\cl{ Queen Mary and Westfield College}
\cl{London E1 4NS}}
\sk
\cl{\tt and}
\sk
\cl{B. Spence
\foot{Address from Oct. 1st 1992: Dept. Physics,
                      University of Melbourne, Victoria 3052 Australia.}}
\sk
{\it\cl{Blackett Laboratory}
\cl{ Imperial College}
\cl{ London SW7 2BZ}}
\sk
\vskip 30pt
\cl {{\bf {Abstract}}\foot
{Talk given at NATO Workshop \lq Low Dimensional Topology and Quantum Field
Theory', Isaac Newton Institute for Mathematical Sciences, Cambridge, Sept.
1992.}}
\vskip 10pt
{\rm
We prove that the covariant and Hamiltonian phase spaces of the
Wess-Zumino-Witten model on the cylinder are diffeomorphic and we derive the
Poisson brackets of the theory.}

\Date{September 1992}

\def\REF{\ref}

\def\gm{{g^{-1}}}

\def\Va{{\delta V V^{-1}}}
\def\Ua{{U^{-1}\delta U}}

\def\half{{1\over2}}

\def\cA{{\cal W}}
\def\pb#1#2{ \{#1,#2\} }
\def\pl{Phys. Lett.\ }
\def\cmp{Commun. Math. Phys.\ }
\def\np{Nucl. Phys.\ }
\def\uup{ \hbox{ \lower3ex\hbox{$\textstyle\otimes$} }}
\def\pbtimes#1#2{ \{{#1}\vbox{\uup\hbox{$\;\;,$}}{#2}\} }
\def\tr {{\rm tr}}

\def\LieG{{\rm Lie}G}
\def\CA {{\cal A}}
\def\kaffa{{\kappa\over4\pi}}
\def\kaape{{\kappa\over8\pi}}
\def\ie{{\it i.e.}}
\def\crr{\cr\noalign{\vskip3pt}}


\sequentialequations


\newsec{Introduction}

  The Wess-Zumino-Witten (WZW) model
\REF\witten{E. Witten, \cmp 92 (1984) 455.} is a two-dimensional non-linear
sigma model with
Wess-Zumino term whose target manifold is a compact, connected Lie group
$G$, and it is a fundamental conformal field theory.
Recently there have been a number of proposals for
deriving a quantum group structure in this model
 \REF\blok{B. Blok, \pl 233B (1989) 359.},
\REF\ash{A.Yu. Alekseev and S. Shatashvili, \cmp 128 (1990) 197; 133
        (1990) 353.},
\REF\faddeev{L.D. Faddeev, \cmp 132 (1990) 131.},
\REF\balog{J. Balog, L. D\c abrowski and L. Feh\'er, \pl 244B (1990) 227.},
\REF\felder{G. Felder, K. Gaw\c edski and A. Kupiainen, \np B299 (1988)
        355; \cmp 117 (1988) 127.},
\REF\gawedski{K. Gaw\c edski, \cmp 139 (1991) 201.},
\REF\bimonte{G. Bimonte, P. Salomonson, A. Simoni and A. Stern, {\it
             Poisson Bracket Algebra for Chiral Group Elements in
              WZNW Model}, preprint UAHEP 9114.},
\REF\chu{M.F. Chu, P. Goddard, I. Halliday, D. Olive and A. Schwimmer,
                \pl B266 (1991) 71.}.
As a consequence of this work,
there has been interest in the phase space structure
of the WZW model \bimonte, \chu,
\REF\us{G. Papadopoulos and B. Spence, {\it The Canonical Structure
            of Wess-Zumino-Witten Models}, preprint Imperial
             91-92/19, QMW 92/2.}. For example,
the authors of Ref. \chu\ have argued that the quantisation of the
Poisson bracket algebra  of a set of variables leads to an exchange algebra.
However, there has been little  study of the {\it global} structure of the
phase
space of the WZW model. This global structure is crucial to the definition
of the symplectic form and the derivation of the Poisson brackets of the
theory. We will discuss this in the following. Some of this work has also been
reported in Ref.
\REF\usb {G. Papadopoulos and B. Spence, {\it The global phase space
structure of the Wess-Zumino-Witten model}, preprint Imperial
             91-92/28, QMW 92/9.}.

There are two approaches to the definition of the phase space of a field
theory.
The first is the Hamiltonian definition.  In this approach, the
space-time is a topological product $M=\Sigma\times \bf{R}$ and the momentum
of the theory is defined as follows:
  \eqn\none{P_i={\partial L(X) \over
\partial\partial_t X^i}.											}
$X$ are the fields and $L$ is the Lagrangian of the theory quadratic in the
time $t$ derivatives of the fields. If there are no constraints in the
theory, the Hamiltonian phase space
 $P_H$ is the co-tangent bundle $T^*Q$ of the configuration space $Q$ of the
model. The Poisson brackets of the theory are
 \eqn\ntwo{\{X^i(x), P_j(y)\}= \delta^i_j(x,y),              }
where $x,y\in \Sigma$.
 These  brackets correspond to the standard symplectic
structure of $T^*Q$. To describe the other definition of the phase space of a
field theory, we start from the Lagrangian $L$ and define the symplectic
current
\eqn\nthree{S^{\mu} = \delta X^i\  \delta { \partial L\over \partial
\partial_{\mu}X^i}. }
This current is conserved subject to the
Lagrangian equations of motion of the theory.  The symplectic form is
\eqn\nfour{\Omega =\int_{\Sigma} S_{\mu}\
d\Sigma^{\mu}.}
																																	$\Omega$ is closed and it does
not depend on the choice of the Cauchy surface $\Sigma$ (see
Ref. \REF\cca {R. Abraham and J.E. Marsden,  {\it Foundations of Mechanics},
       Benjamin/Cummings  Publishing  Company, 1978.} for
discussions of the Lagrangian approach to the phase space).  If the theory does
not have constraints, it is possible to relate the Lagrangian and Hamiltonian
definition of the symplectic form by a Legendre transform.  One way to
parameterise the symplectic form $\Omega$ is in terms of the initial data
$f(x) = X(x,0)$ and $w(x) = (\partial_t X)(x,0)$.  However, if the solutions
of the Lagrangian equations of motion are known, then it is possible to
parameterise $\Omega$ in terms of the parameters of the solutions of the
theory.  The space of solutions of the Lagrangian equations of  motion of a
theory  equipped with the symplectic form of eqn \nfour\  is called
the covariant phase space $P_C$  (see
Ref. \REF\ccb {G. Zuckerman, in
         {\it Mathematical Aspects of String Theory}, ed. S.-T. Yau,
            World Scientific, Singapore 1987;\hfill\break
            \v C. Crnkovi\'c and E. Witten, in {\it Three Hundred Years of
           Gravitation}, eds. S.W. Hawking and W. Israel, C.U.P.
           Cambridge, 1987.} for recent discussions).  The
covariant phase space is particularily suited to the study of field
theories where we know the space of  classical solutions, such as in the case
of
the WZW model.

    In the following, we will
give a new parameterisation of the space of solutions of the WZW
model, and use this to prove that the
Hamiltonian and covariant phase spaces are diffeomorphic.
The
Poisson brackets of the theory will also be derived.
Finally, we will compare our approach with other formulations that have
appeared in the literature.


\newsec{The WZW Model}

In the Hamiltonian approach, one may consider the WZW model as a
two-dimensional non-linear sigma model with Wess-Zumino term,
whose target space is a
group manifold $G$. Applying the usual Hamiltonian analysis to
this sigma model action, one finds directly that
the phase space $P_H$ of the WZW model is the co-tangent
bundle of its configuration space $LG$, \ie\ $P_H=T^*LG$ (see Ref.
\REF\hwzw{I. Bakas and D. McMullan, \pl B189 (1987) 141.}).

Now we consider the covariant approach to the phase space.  The symplectic
form of the WZW model is (see Ref. \chu, for example)
\eqn\naseven{    \Omega = -\kaape\,
   \int_0^1\!\!dx\,\tr\,\Big((\gm\delta g)\partial_+(\gm\delta g)
                 - (\delta g\,\gm)\partial_-(\delta g\,\gm)\Big),}
     where $\kappa$ is the coupling constant of the model, $(x,t)$,
$0\leq x <1, -\infty <t <+\infty$
are the co-ordinates of the cylinder $S^1\times
{\bf{R}}$ and $x^{\pm}=x\pm t$.  This symplectic form is closed and time
independent (we take $t=0$ in the following).

One way to parameterise  $\Omega$ is in terms of the initial conditions
$f(x) = g(x,0)$ and $w(x) = (g^{-1} \partial_t g)(x,0)$ on the Cauchy
surface $t=0$.  In terms of the functions  $f$ and $w$, the symplectic form
$\Omega$ becomes
\eqn\naeight{\eqalign{\Omega=-\kaape\,
   \int_0^1\!\!dx\,\tr\,\Big( &\half f^{-1}\delta
f\partial_x(f^{-1}\delta f)  - \half\delta f
f^{-1}\partial_x(\delta f f^{-1})
\cr &
 +(f^{-1}\delta f)^2 w + f^{-1}\delta
f\delta w \Big).}}

To construct the covariant phase space $P_C$, we should introduce a
parameterisation of the space of solutions of the WZW model.  The equations
of motion are
\eqn\nafive{\partial_-(\partial_+  g\ g^{-1})=0. }
The equations of motion are invariant under the semi-local transformations
$g\ \rightarrow l(x^+)\  g \  r^{-1}(x^-)$
and the corresponding currents are
\eqn\nasix{
J_+= -\kaffa\partial_+g g^{-1},  \qquad J_-= \kaffa g^{-1}
                          \partial_-g. }
There are
several suggestions in the literature as to  how to parameterise the space
of solutions of the WZW model.  In the following section we will discuss
the parameterisation of Ref. \usb,
and in the last section we will compare
this parameterisation  with  others  in the literature.


\newsec{The Poisson brackets}

In Ref. \usb, we have parameterised  the space of
solutions to the field equations of the WZW model as follows
   	\eqn\bthree { \eqalign{   g(x,t) &= U(x^+) \cA(A;x^+,x^-) V(x^-), \crr
                 \cA(A;x^+,x^-) &=
                     P\exp\!\int_{x^-}^{x^+}\!A(s)ds,\cr }}
where $U$ and $V$ are periodic maps from the real line ${\bf{R}}$ to the
group $G$, and the field $A$ in the path-ordered exponential is a
 $(\LieG)^*$-valued periodic one-form on
the real line.   The expression for $g(x,t)$ in Eqn. \bthree\
is then periodic in $x$ and solves
the field equations \nafive.  To prove the periodicity of $g$, it is enough
to show that ${\cA}(A;x^++1,x^-+1)={\cA}(A;x^+,x^-)$.  One way to verify this
is to use the power series expansion of ${\cA}$ and change variables.  This
gives ${\cA}(A;x^++1,x^-+1)={\cA}({\hat {A}};x^+,x^-)$ where
${\hat{A}}(x)=A(x+1)$. Using the periodicity of $A$ we can then prove that
$g$ is periodic.   An alternative way is to use the formula
${\cA}(A;x^++1,x^-+1)=m(x^+){\cA}(A;x^+,x^-) m^{-1}(x^-)$$={\cA}(A^m;x^+,x^-)$,
where $m(x)={\cA}(A;x+1,x)$ and $A^m=\partial_xm(x) m^{-1}(x) + m(x) A(x)
m^{-1}(x)$.  However $\partial_x m(x)=A(x+1) m(x)- m(x) A(x)$, and from the
periodicity of $A$ we get $A^m=A$.  This again proves the periodicity of
$g$.

Next,
using the  parallel transport equation
      	\eqn\bfour{  \partial_s\cA (A;s,x^-) = A(s)\ \cA (A;s,x^-),}
we can prove that $g$ in eqn. \bthree\ satisfies the equations of motion of
the WZW model. Choosing a point $x_0$ on the real line, we can write the
 solution given in eqn. \bthree\  in a chirally factorised form
$g(x,t)=u(x^+, x_0)v(x^-, x_0)$, by using the identity
 ${\cA}(A;x^+,x^-)={\cA}(A;x^+,x_0){\cA}(A;x_0,x^-)$.
However, this  factorisation  depends on the
choice of the point $x_0$.

Inserting the solution \bthree\ into the symplectic form \naseven\ gives
 \eqn\bfive{  \eqalign{
    \Omega = -\kaape\,\int_0^1\!dx\,\tr\,\Big( &(\Ua)\partial_x(\Ua)
              + 2(\Ua)^2\,A + 2(\Ua)\delta A \cr &
      - (\Va)\partial_x(\Va) -2(\Va)^2\,A
+ 2(\Va)\delta A \Big).  \cr}  }
The solution $g$ of the WZW equations of motion given in the
parameterisation \bthree\ is invariant under the
transformations
\eqn\bsix{\eqalign{
      U(x)&\rightarrow U(x) h(x),
         \quad V(x)\rightarrow h^{-1}(x) V(x),
\cr
A(x) &\rightarrow
-h^{-1}(x)\partial_x h(x)+ h^{-1}(x)A(x)\,h(x),} }
where $h\in LG$.
To prove this, we observe that under these transformations
$\cA(A;x^+,x^-)\rightarrow h^{-1}(x^+) \cA(A;x^+,x^-) h(x^-)$.
The phase space $P_C$ of the WZW model is then the space of fields
$\{U,V,A\}$, modulo the  transformations \bsix.
This is ${LG\times LG\times
{\CA} \over LG}$ where $\CA$ is the space of $G$-connections over the
circle.
\foot{There two ways to think about the field $A$.  The first is as a
periodic one-form on the real line, valued in $(\LieG)^*$.
The second is as a connection over the
circle $S^1$.  We have identified $\LieG$ with $(\LieG)^*$ using the invariant
metric.}
This is diffeomorphic to $T^*LG$, \ie\ it is the same as the phase
space $P_H$ derived from the Hamiltonian treatment of the theory.

The symplectic form \bfive\ is degenerate along the directions of
the action \bsix\ of the loop group $LG$.
We may deal with this by gauge-fixing or by enhancing the phase
space of the theory and then imposing constraints \usb.  In the following,
we will use the gauge-fixing method.  We may choose as a gauge fixing
condition  $U = e$ where $e$ is the identity element of the loop group $LG$.
This is a good gauge choice, as $LG$ acts freely and transitively on the
space of $U$'s, $\{U\}$. The symplectic form \bfive\ then becomes
 \eqn\beight{
    \Omega = -\kaape\,\int_0^1\!dx\, \tr\,\Big( -(\Va)\partial_x(\Va)
            - 2(\Va)^2\,A + 2(\Va)\delta A
         \Big).  }
This symplectic form is not degenerate and is invertible.\foot{Notice that
if we set $A = \half(\partial_x f f^{-1} + fwf^{-1})$ and
$V = f$ in the symplectic form \naeight\ then we get the symplectic
form \beight.}

 The simplest way to invert the form \beight\ is to first rewrite it in terms
of a local parameterisation $X^i(x)$ for the maps $V$ ($V=V(X)$). This gives
 \eqn\bnine{
   \Omega = -\kaape\int_0^1\!dx\Big( -(R^a_i\delta X^i)\partial_x(R^a_j\delta
             X^j) - {f_{ab}}^c R^a_iR^b_jA_c\,\delta X^i\ \delta X^j
      + 2 R^a_i\, \delta X^i\delta A_a\Big),  }
where $\delta VV^{-1}=R^a t_a$.
The remarkable feature of this expression for the
form $\Omega$ is that one does not
need to invert any differential operator in order to invert the form
({\it c.f.} Refs. \chu,\us, where in order to invert the symplectic
form it was necessary to find the inverse of
the operator $\partial_x$ on the circle).
The gauge
$U=e$ parameterises the symplectic form on $T^*LG$ in terms of the right
trivialisation of $T^*LG$ and the gauge $V=e$ parameterises the same
symplectic form in terms of the left trivialisation. The
inversion of the form \bnine\ is straightforwardly carried out, and
leads to the Poisson brackets ($\beta = -{4\pi\over\kappa}$)
\eqn\bten{\eqalign{ \pb{X^i(x)}{X^j(y)} &= 0,\crr
                   \pb{X^i(x)}{A_a(y)}
                  & =\beta R^i_aX(x) \delta(x,y), \crr
       \pb{A_a(x)}{A_b(y)} &= \beta\Big(\delta_{ab}\partial_x +
                     {f_{ab}}^c A_c(x) \Big)\delta(x,y), \cr}
              }
where $\delta(x,y)$ is the delta function on $S^1$.

Using Eqn. \bten, we can calculate Poisson brackets involving $V$ and
$A$ -- for example $\pbtimes{V(x)}{V(y)} = 0$, $\pb{V(x)}{A_a(y)} =\beta
 V(x)t_a\delta(x,y)$. In this gauge, the WZW currents \nasix\ become
$J_+ = -{\kappa \over 4\pi} A$, $J_- = {\kappa \over 4\pi}
(V^{-1}\partial_xV - V^{-1}AV) $, and it can be verified by a
straightforward calculation that their Poisson bracket algebra is isomorphic
to two commuting copies of a Kac-Moody algebra with a central extension.


\newsec{Discussion}

We would now like to discuss how our results relate to other
work in the literature.
In Ref. \witten, Witten proposed that  the  general solution of the WZW
equations of motion is $g(x,t)=U(x^+) V(x^-)$
where $U$ and $V$ are maps from the
circle into the group $G$.  In this parameterisation the covariant phase
space is $LG\times LG$ where $LG$ is the loop group of the group $G$.
Recently, another parameterisation of the space of solutions of the WZW
model was proposed by Chu {\it et al} in Ref. \chu.  In this parameterisation
the solution of the WZW model factorises as $g(x,t)=u(x^+) v(x^-)$.
However, the functions $u$ and $v$ are not periodic and satisfy the
conditions $u(x+1)=u(x) M$, $v(x+1)=M^{-1} v(x)$, where $M\in G$ is the
monodromy. In addition, the authors of Ref. \chu\ observed that
the solution $g=uv$ of the WZW model remains invariant under the action
$u(x)\rightarrow u(x) k$, $v(x)\rightarrow k^{-1} v(x)$, $M\rightarrow
k^{-1} M k$ of the group $G$, with $k\in G$.   An  equivalent way
to write the solution $g$ of the WZW model in this parameterisation is
$g(x,t)=U(x^+) M^{2t} V(x^-)$, where $U,V$ are periodic.  The covariant phase
space of the WZW model in the parameterisation of Ref. \chu\ is ${LG\times
LG\times G \over G}$.  Neither this covariant phase space, nor that of
Ref. \witten, are
diffeomorphic to the Hamiltonian phase space $T^*LG$ of the WZW model -- for
example, $\pi_2(LG\times LG) = \pi_2( {LG\times LG\times G \over G}) = {\bf
Z}\oplus{\bf Z}$, whereas $\pi_2(T^*LG) = {\bf Z}$ (for $G$ simple and simply
connected).

To relate the parameterisation of the space of solutions of the WZW model
of Ref. \chu\ to the parameterisation of Section Three, the symmetries of the
space of solutions  of the  WZW model (Eqn. \bsix) can be treated by
choosing  the gauge-fixing conditions to be different from those considered
in Section  Three above.  For example in the parameterisation \bthree\ one
can gauge-fix the connection $A$ so that it is a {\it constant}  connection
over the circle.  The residual  transformations for this gauge-fixing are the
constant gauge transformations.  The constant gauge transformations are
parameterised by the elements of the group $G$ and they act on the
parameters of the solutions as  $U\rightarrow U k$, $V\rightarrow k^{-1} V $
and $A\rightarrow k^{-1}A k$ where $k\in G$ and $A$ is a constant
connection.  This parameterisation is the one given in Ref. \chu. The
resulting phase space of the theory is
${LG\times LG \times G\over G}$, if we parameterise the solutions in terms
of the monodromy $M$ of the connection $A$.
 The reason that this phase space is different from $T^*LG$
is that there is a
Gribov-Singer ambiguity associated with this gauge fixing; note that this is
the
one-dimensional analogue   of the four-dimensional Yang-Mills Gribov-Singer
ambiguity
\REF\gribov{V.N. Gribov, \np B139 (1978) 1;
             \hfill\break I. Singer, \cmp 60 (1978) 7.}.
The $k$-symmetry just mentioned can be further gauge
fixed by choosing $A$ to be in the Cartan  subalgebra $\bf{h}$ of $\LieG$,
and this parameterisation was used to calculate the Poisson brackets of this
theory in Refs. \chu, \us.  Finally, the parameterisation of Ref. \witten\
corresponds to the gauge-fixing choice $A=0$. However, not all connections
can be brought into this form by a gauge transformation.

An alternative way to compare the different parameterisations of the space of
solutions of the WZW model is to study the initial values  $f(x)=g(x,0)$ and
$w(x)=(g^{-1}\partial_t g)(x,0)$ of the field $g$ that correspond to these
parameterisations.  In general $f$ and $w$ are independent functions.  It is
easy to see that the parameterisation of eqn.  \bthree\  corresponds to the
most
general Cauchy data of the theory.  However, the parameterisations of Refs.
\chu\ and \witten\ are associated to a {\it subset} of the available Cauchy
data
of this theory.  The restriction that the parameterisation of
Ref. \chu\ imposes on the Cauchy data is that the holonomy of the
connection ${\hat{w}} =
\half (f^{-1}\partial_x f- w)$ does not depend on  the point
chosen to evaluate it, i.e. $\partial_x{\cA}(\hat{w};x+1,x)=0$. The
constraint that the parameterisation of Ref.
 \witten\ imposes on the Cauchy data
is that the holonomy of the connection ${\hat {w}}$ on the circle $S^1$ must be
the
identity group.

In conclusion, the parameterisation of the solutions of the WZW model given
in Section Three (Eqn. \bthree) is general in the sense that it is invariant
under a larger symmetry than other parameterisations considered
in the literature, and the latter can be thought of as locally-valid
gauge-fixed versions of it.
In our parameterisation, the covariant canonical phase space of the WZW model
is diffeomorphic to the Hamiltonian phase space of the theory, and the
calculation
of the Poisson brackets is straightforward.

\noindent{\bf Acknowledgements:} This work was supported by the SERC.

\listrefs

\bye